\documentclass[twocolumn]{aastex63}
\usepackage[utf8]{inputenc}
\usepackage{graphicx}
\usepackage{amsmath}
\usepackage{hyperref}
\graphicspath{{Figures/}}

\newcommand{\dpart}[2]{\frac{\partial #1}{\partial #2}}
\newcommand{\deriv}[2]{\frac{d#1}{d#2}}
\newcommand{\gcc}{~{\rm g\,cm^{-3}}}

\begin{document}
\title{Removal of Hot Saturns in Mass-Radius Plane by Runaway Mass Loss}
\author[0000-0002-5113-8558]{Daniel P. Thorngren}
\affil{Trottier Institute for Research on Exoplanets (iREx), Universit\'e de Montr\'eal, Quebec, Canada}
\affil{Department of Physics \& Astronomy, Johns Hopkins University, Baltimore, MD, USA}
\author[0000-0002-1228-9820]{Eve J.~Lee}
\affil{Department of Physics and Trottier Space Institute, McGill University, Montr\'eal, Qu\'ebec, H3A 2T8, Canada}
\affil{Institute for Research on Exoplanets (iREx), Universit\'e de Montr\'eal, Quebec, Canada}
\author[0000-0002-7727-4603]{Eric D. Lopez}
\affil{NASA Goddard Space Flight Center, 8800 Greenbelt Rd, Greenbelt, MD 20771, USA}
\affil{GSFC Sellers Exoplanet Environments Collaboration}

\begin{abstract}
    The hot Saturn population exhibits a boundary in mass-radius space, such that no planets are observed at a density less than $\sim 0.1~{\rm g\,cm^{-3}}$.  Yet, planet interior structure models can readily construct such objects as the natural result of radius inflation.  Here, we investigate the role XUV-driven mass-loss plays in sculpting the density boundary by constructing interior structure models that include radius inflation, photoevaporative mass loss and a simple prescription of Roche lobe overflow.  We demonstrate that planets puffier than $\sim 0.1~{\rm g\,cm^{-3}}$ experience a runaway mass loss caused by adiabatic radius expansion as the gas layer is stripped away, providing a good explanation of the observed edge in mass-radius space.  The process is also visible in the radius-period and mass-period spaces, though smaller, high-bulk-metallicity planets can still survive at short periods, preserving a partial record of the population distribution at formation.
\end{abstract}

\section{Introduction} \label{sec:introduction}
Mass loss processes transform large planets into their smaller counterparts and leave imprints in the observed exoplanet population in the shape of deficits. The most famous example is the radius gap that separates the gas-enveloped mini-Neptunes ($\sim$2--4$R_\oplus$) from the more rocky super-Earths ($\sim$1--1.7$R_\oplus$) which was predicted to exist by theories of photoevaporative mass loss \citep[e.g.][]{Lopez2012,Owen2013} and later confirmed observationally when the host star samples were confined to those that were well-characterized by high-resolution spectra \citep[e.g.][]{Fulton2017,Fulton2018}, asteroseismology \citep[e.g.][]{VanEylen2018}, or followed up by Gaia and detailed statistical analysis \citep[e.g.][]{Hsu2019}.%
\footnote{The same gap is expected to appear prior to mass loss processes purely from the physics of gas accretion \citep[e.g.][]{Lee2021, Lee2022a}, while subsequent mass loss such as photoevaporation or core-powered envelope mass loss \citep[e.g.][]{Ginzburg2018,Gupta2019,Gupta2020} further tune the planet population.  Alternatively, the radius gap may not have anything to do with gas envelope and instead stem from two distinct core compositions (e.g. \citealt{Zeng2019}, but see \citealt{Aguichine2021}).}

Another deficit we see in exoplanet population is the sub-Jovian desert, a roughly triangular region in mass-period and radius-period space where we see no Saturn size objects inside orbital periods of a few days, first reported in mass space by \cite{szabo2011} then in radius space by \cite{Beauge2013} and then later confirmed to exist in both mass-period and radius-period spaces by \citet{Mazeh2016}. This sub-Jovian desert subsumes the so-called ``photoevaporation desert'' which refers to the lack of short-period mini-Neptunes whose incident flux would correspond to $\sim$650 times that received by the Earth \citep[e.g.][]{Lundkvist2016}, suggesting photoevaporative mass loss to be at play in sculpting at least part of the sub-Jovian desert. Yet another way to view the desert is to consider the orbital period distribution of sub-Saturns (4--8$R_\oplus$). Using a hydrodynamic mass loss model of \citet{Kubyshkina2018}, \citet{Hallatt2022} demonstrated that mass loss alone could reproduce the fall-off in the sub-Saturn occurrence rate at short periods with the stripped Saturns accounting for a fraction of the observed mini-Neptunes and super-Earths.

Although photoevaporation is expected to be an efficient mechanism to strip small planets ($\lesssim 8 R_\oplus$) of their gaseous envelopes, detailed radiative-hydrodynamic calculations have found that gas giants are resilient against mass loss, losing less than 1\% of its total mass over the star's main-sequence lifetime \citep[e.g.][]{Murray-Clay2009,Owen2012}. It follows that although the lower boundary of the sub-Jovian desert is likely shaped by photoevaporative mass loss, the upper boundary of the desert may require a different explanation. 

\cite{Valsecchi2014} proposed that hot Jupiters that are excited onto high enough eccentricities by e.g., Kozai can undergo catastrophic Roche lobe overflow and lose all their envelopes, transforming into rocky super-Earths at ultra short periods. While some of the short-period rocky planets may be stripped giants, the lack of strong correlation between the occurrence of ultra-short period planets and the host star metallicity (as opposed to the strong correlation between the occurrence of hot Jupiters and high metallicity of the host star) suggest that this is likely not the main origin channel \citep{Winn2018}. 

Instead of mass loss, \cite{Matsakos2016} considered the boundaries of the desert to be traced by the history of dynamical migration. They proposed that both the upper and the lower boundaries can be reproduced by planets arriving at their closest in orbits by high-eccentricity migration with the final orbital distance set by either the Roche lobe radius or to twice the pericenter distance (of the initially excited eccentric orbit) as expected from secular chaos \citep{Wu2011}. The different shape of the upper and the lower boundary (in the mass-period space) reflects the different mass-radius relationship followed by smaller vs. larger planets. \citet{OwenLai2018} combined the expectation from high-eccentricity migration (along with tidal decay) with photoevaporative mass loss and concluded that both mechanisms are required to explain the overall shape with the former being responsible for the upper boundary and the latter being responsible for the lower boundary.

\begin{figure}
    \centering
    \includegraphics[width=\columnwidth]{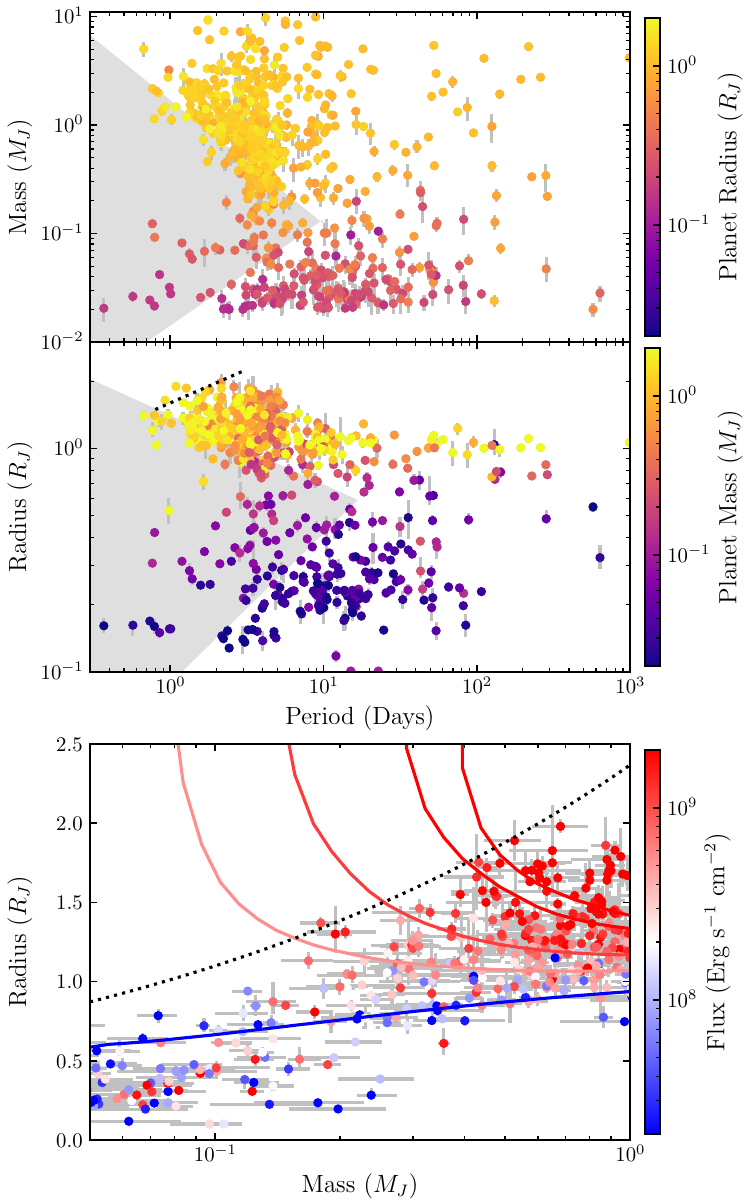}
    \caption{Three views of the observed exoplanet population (our selection criteria are described in Section \ref{sec:methods}).  The top two panels show the conventional view (mass/radius vs. period) of the sub-Jovian desert, which is the shaded area \citep[following][]{Mazeh2016}.  We are primarily interested in the low-density boundary seen in mass-radius space (constant $0.1 \gcc$, black dotted line, bottom panel), which is roughly equivalent to the black dotted line in the middle panel and partially overlaps with the mass-period desert.  We have also plotted model radius lines (see Section \ref{sec:methods}) for various incident fluxes demonstrating that planets with density less than  $0.1 \gcc$ can exist but nevertheless do not in nature.}
    \label{fig:overview}
\end{figure}

All aforementioned literature relied on their assumed mass-radius relationship for different categories of planets. In this letter, we investigate directly the sub-Jovian population in the radius-mass space and self-consistently compute the evolution of planetary interiors under mass loss. We are motivated by a clean upper boundary in the radius-mass plane that is well-approximated by a constant $\sim 0.1\gcc$ (see the bottom panel of Figure \ref{fig:overview}) which also coincides with the region of the parameter space where highly irradiated planets become very large and the mass-radius relationship steepens \citep{Thorngren2018}, suggestive of photoevaporative mass loss playing a major role in creating the boundary. By comparison, this density boundary manifests as the very upper edge of the hot Jupiter population in the radius-period space (middle panel) while being coincident with the usual sub-Jovian desert in the mass-period space (top panel).  

This paper is organized as follows. In Section \ref{sec:methods}, we describe how we build time-evolving models of interior structure accounting for XUV-driven mass loss. Our results are presented in Section \ref{sec:results} and we conclude in Section \ref{sec:disc}.

\section{Methods} \label{sec:methods}
We start with gathering data on the observed exoplanets, which we collect from the NASA Exoplanet Archive (DOI: \href{https://catcopy.ipac.caltech.edu/dois/doi.php?id=10.26133/NEA12}{10.26133/NEA12}; \citet{Akeson2013}) and Exoplanet.eu \citep{Schneider2011}.  We combine this data and select confirmed planets with radial-velocity mass measurements more precise than 30\% and radius measurements more precise than 20\%. Additionally, we exclude Kepler-13b, Kepler-435b, Kepler-87b, and KELT-9b as their quoted stellar properties between listed references are inconsistent or their quoted flux is so high that they cannot be self-consistently described within our model framework.  While we are primarily interested in planets between 0.1 and $1 M_J$ and with $a < 0.1$ AU, we include planets outside these ranges for context.  It is not our goal to reproduce the absolute abundance of planets and so we do not correct for observational biases.  Rather, what is relevant is whether our model is able to account for the {\it paucity} of planets in spaces where they are favorable to be observed (i.e., large radii and short period).

The majority of our interior structure evolution model is the same as was used in \cite{Thorngren2019}.  These are 1-dimensional models constructed by solving the equations of hydrostatic equilibrium, mass conservation, and an equation of state (EOS):
\begin{gather}
    \frac{\partial P}{\partial m} = -\frac{G m}{4 \pi r^4},\\
    \frac{\partial r}{\partial m} = \frac{1}{4 \pi r^2 \rho},\\
    \rho = \rho(P,T),
\end{gather}
where $m$ and $r$ are the enclosed mass and radii of each layer respectively, $P$ is pressure, $\rho$ is density, $T$ is temperature, and $G$ is the gravitational constant.  We use the \cite{Chabrier2019} EOS for a solar-ratio mixture of hydrogen and helium, and a 50-50 mixture of rock and ice for the metals \citep{Thompson1990}.

To set the bulk planet metallicity, we use the mass-metallicity distribution from \cite{Thorngren2016}, in which the logarithm of the metal mass $M_z$ is normally distributed with a median $M_z = 0.182 M^{0.61}$.  This median value is our default; in Section \ref{sec:results}, we experiment with varying the bulk metallicity.  The offset from the mean (in log-metallicity) is parameterized by $Z_{zpl}$ the z-score (number of sigma away from the mean) -- e.g. $Z_{zpl}=1$ selects the 84th percentile.  This is the same approach as was used in \citep{Thorngren2019}.  This is equivalent to multiplying the median by a factor of $10^{\sigma Z_{zpl}}$, where $\sigma=0.26$. The advantage of this approach is that drawing $Z_{zpl}$ from a standard normal distribution correctly reproduces the mass-metallicity dispersion.  The core mass is set to half the metal mass, but with a minimum of 5 $M_\oplus$ and a maximum of 15 $M_\oplus$; any remaining metals are mixed into the envelope.

Modelling the thermal evolution of the planet is slightly more complex than in past cases because the mass loss process can also affect the thermal state of the planet.  Our model tracks the thermal state of the planet in terms of the envelope specific entropy $s$, so we seek to calculate its rate of change with time in order to integrate.  We start with conservation of total energy $E$, split into the change in total energy in the core $E_c$ and the change in total energy in the envelope $E_e$:
\begin{equation}
    \deriv{E}{t} = \deriv{E_c}{t} + \deriv{E_e}{t},
    \label{eq:dEdt}
\end{equation}
where $t$ is time.  We assume an isothermal core, and that the core can release energy into the envelope rapidly enough that the core-envelope boundary has no temperature jump.  We adopt $c = 7.5 \times 10^6$ erg/g/K \citep[chosen as a typical value for rock, see][]{Waples2004} for the specific heat capacity of the core.  The core temperature $T_c$ can then change either because the envelope specific entropy changes or because the envelope loses mass and the core-envelope boundary intersects at a lower pressure and therefore temperature (even at constant entropy).  We find that this core depressurization heat is a negligible effect, but we include it for completeness.  The energy change in the envelope is readily calculated from the change in entropy \citep[see e.g.][]{Thorngren2016}.  Putting this all together and solving for the change in entropy with time, we have
\begin{gather}
    \deriv{E}{t} = c M_c \left( \dpart{T_c}{M} \dpart{M}{t} + \dpart{T_c}{s} \dpart{s}{t} \right) + \deriv{E_e}{s} \deriv{s}{t}; \\
    \deriv{s}{t} = \left. \left( \deriv{E}{t} - c M_c \dpart{T_c}{M} \dpart{M}{t} \right) \middle/ \left( \dpart{E_e}{s} + c M_c \dpart{T_c}{s} \right), \right. \label{eq:dsdt}
\end{gather}
where $M$ is the total mass of the planet, and $M_c$ is the mass of the core.  These derivatives are numerically computed by constructing and evaluating the differences between two static models at slightly different envelope masses and at slightly different specific entropy for $\partial T_c / \partial M$ and $\partial T_c / \partial s$, respectively. The change in envelope energy with entropy is calculated by integrating $\partial s / \partial E = 1/T(P,s)$ with respect to mass across the envelope \citep[as in e.g.][]{Lopez2012} with $T(P,s)$ supplied by the adopted equation of state.

We now consider the energy sources and sinks and equate them Eq.~\ref{eq:dEdt}:
\begin{equation}
    \deriv{E}{t} = L_{\rm rad} + \pi R^2 (\epsilon F - 4F_{\rm int})
\end{equation}
where $L_{\rm rad}$ is radioactive heating, $R$ is the radius of the planet, $\epsilon$ is the anomalous heating efficiency for which we adopt the median inferred value from \cite{Thorngren2018}, $F$ is irradiation flux, and $F_{\rm int}$ is the internal cooling flux of the planet.  Radioactive heating is computed from the decay of U235, U238, Th232, and K40 following \cite{Nettelmann2011} and \cite{Lopez2012}, using \cite{Anders1989} for the radioactive elemental abundances relative to silicon.  We find that $L_{\rm rad}$ is an extremely minor component of the energy balance.  Finally, $F_{\rm int}$ is calculated for a given envelope specific entropy, surface gravity, and insolation using the non-grey atmosphere models of \cite{Fortney2007}.

The anomalous heating is a key feature of our models. Their inclusion is motivated by the fact that hot Saturns are above the $T_{eq}=1000$ K inflation threshold \citep{Demory2011,Miller2011} and that many of them are too large to be explained by a low metallicity alone \citep[$R>1.2R_J$, see][]{Thorngren2016}.  Our adoption of \citet{Thorngren2018} inflation power fits---which were based on planets with $M > 0.5 M_J$---is reasonable because most of the planets below that regime fit the models quite well; the cut was only made to avoid the influence of a few dense outliers such as HD 149026 b \citep[see e.g.][]{Sato2005,Fortney2006}.\footnote{Even completely eliminating anomalous heating is not sufficient to explain their small size, so \citet{Thorngren2018} posited that such planets are simply very metal rich, either from their formation or due to mass-loss.}

To model mass-loss, we use the following expression:

\begin{gather} \label{eq:massLoss}
\dot{M} \approx \frac{\eta \pi F_{XUV} R^3_{XUV}}{G M K_t} 
    \approx \frac{3}{4} \frac{\eta F_{XUV}} {G K_t \rho_{\rm XUV}};\\
K_t = 1 - \frac{3}{2\xi} + \frac{1}{2\xi^3}; \\
\xi = \frac{R_{hill}}{R_{XUV}},
\end{gather}
where $\eta$ is the mass-loss efficiency factor, $K_t$ is a tidal correction factor, $R_{\rm hill}$ is the Hill radius, $R_{XUV}$ is the radius below which the planet becomes opaque to XUV irradiation, which we set at 10 nanobars (a conservative limit of \citealt{Lopez2012}), and $F_{XUV}$ is the extreme ultraviolet flux from the parent star which we interpolate from the stellar evolution grid of \cite{Johnstone2021} using the median stellar spin.  While Eq.~\ref{eq:massLoss} mirrors that of energy-limited approximation from \citet{Watson1981, Lopez2012, Lopez2013}, we do not assume energy-limited mass loss.  Instead we use the results of \citet{Caldiroli2022}, who apply an empirical fitting to a detailed hydrodynamics code ATES \citep{Caldiroli2021} that encompasses both energy-limited and non-energy-limited regime of mass loss, reported as an effective mass-loss efficiency $\eta$ that varies non-trivially with incident flux and planetary potential.  We find that hot Saturns fall in their low gravity regime where $\eta$ generally ranges in the few tens of percent.  The deposition depth of XUV irradiation is another important parameter for XUV-driven mass-loss models; it is often quoted as a few nanobars \citep{Murray-Clay2009,Caldiroli2021} for giants, although they may be deeper in depending on the mass of the planet \citep{OwenLai2018, Ionov2018}.  We do not need to set this parameter, as ATES distributes the energy according to the ionization cross-sections in the model atmosphere.  Finally, we have rewritten the mass loss rate $\dot{M}$ in terms of $\rho_{\rm XUV}$ the planet bulk density using the XUV radius to highlight the connection between the observed density boundary for hot Saturns (Figure \ref{fig:overview}) and XUV-driven mass loss.

For planets that grow to overflow their Roche lobes, we model both the mass loss from the overflow as well as the effect on the planets' semimajor axes, which expand in proportion to the mass lost from the conservation of angular momentum \citep{Valsecchi2014}:
 \begin{equation}
     \frac{\dot a}{a} = -2 \frac{\dot M}{M} \left( 1 - \frac{M}{M_*}\right).
 \end{equation}
 Here, $a$ is the planet's semimajor axis and $\dot a$ is the rate of change thereof, and $M_*$ is the mass of the parent star.  During Roche lobe overflow, we assign the mass rate of change $\dot M$ to an arbitrary exponential decay with a half-life of 13.8 Myr.  We choose this scheme instead of directly modeling the detailed physics of Roche lobe overflow \citep[e.g.,][]{Valsecchi2014} since the true rate of Roche lobe overflow would be fast compared to the lifetime of planets and their thermal evolution.  In fact, the net effect of Roche lobe overflow in our models is to cause the planet to rapidly lose as much mass as is necessary to expand the orbit to the point where it is no longer overflowing.  This leads to an interesting interplay with XUV-driven mass loss which we will discuss in the next section.

\begin{figure}
    \centering
    \includegraphics[width=\columnwidth]{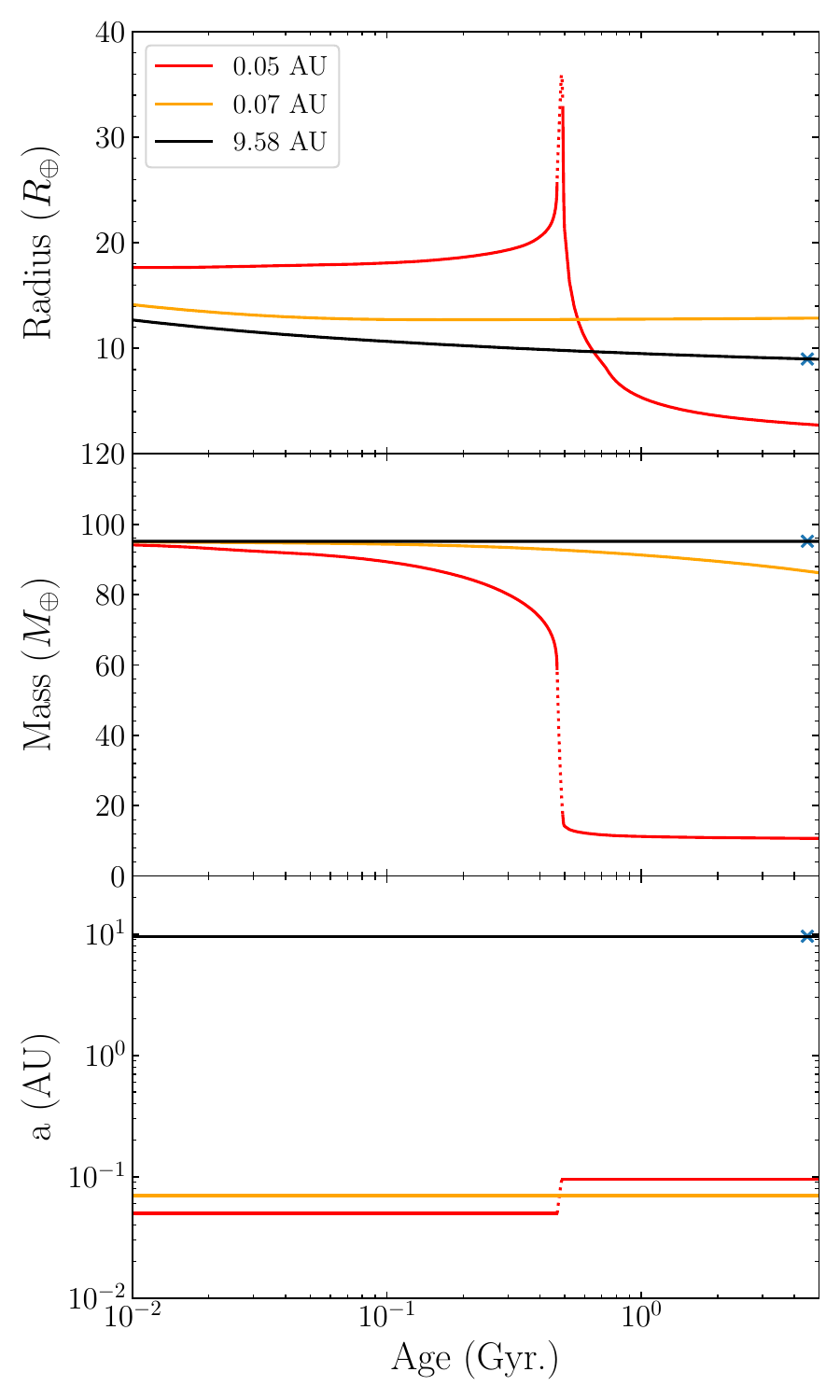}
    \caption{The thermal and mass evolution of a Saturn-mass planet at various semimajor axes.  The planet bulk metallicity was tuned to $Z_p=.225$ to match Saturn's radius (blue x) at 4.5 Gyr and 9.58 AU.  If instead placed at 0.07 AU, the planet experiences modest mass-loss over Gyr timescales, but the radius is mostly unchanged.  At 0.05 AU, the planet loses mass at increasing rates until the runaway mass-loss strips the envelope from the planet.  For part of this process the planet is also filling its Roche lobe (dotted line), ejecting about 40 $M_\oplus$ and expanding the semimajor axis out to .095 AU.  Even after Roche lob overflow fully stops (the overflow appears intermittent, see \citep{Valsecchi2014} and Sec. \ref{sec:results}), XUV-driven mass-loss continues to erode what little is left of the H/He envelope.}
    \label{fig:exampleTracks}
\end{figure}

\begin{figure*}
    \centering
    \includegraphics[width=\textwidth]{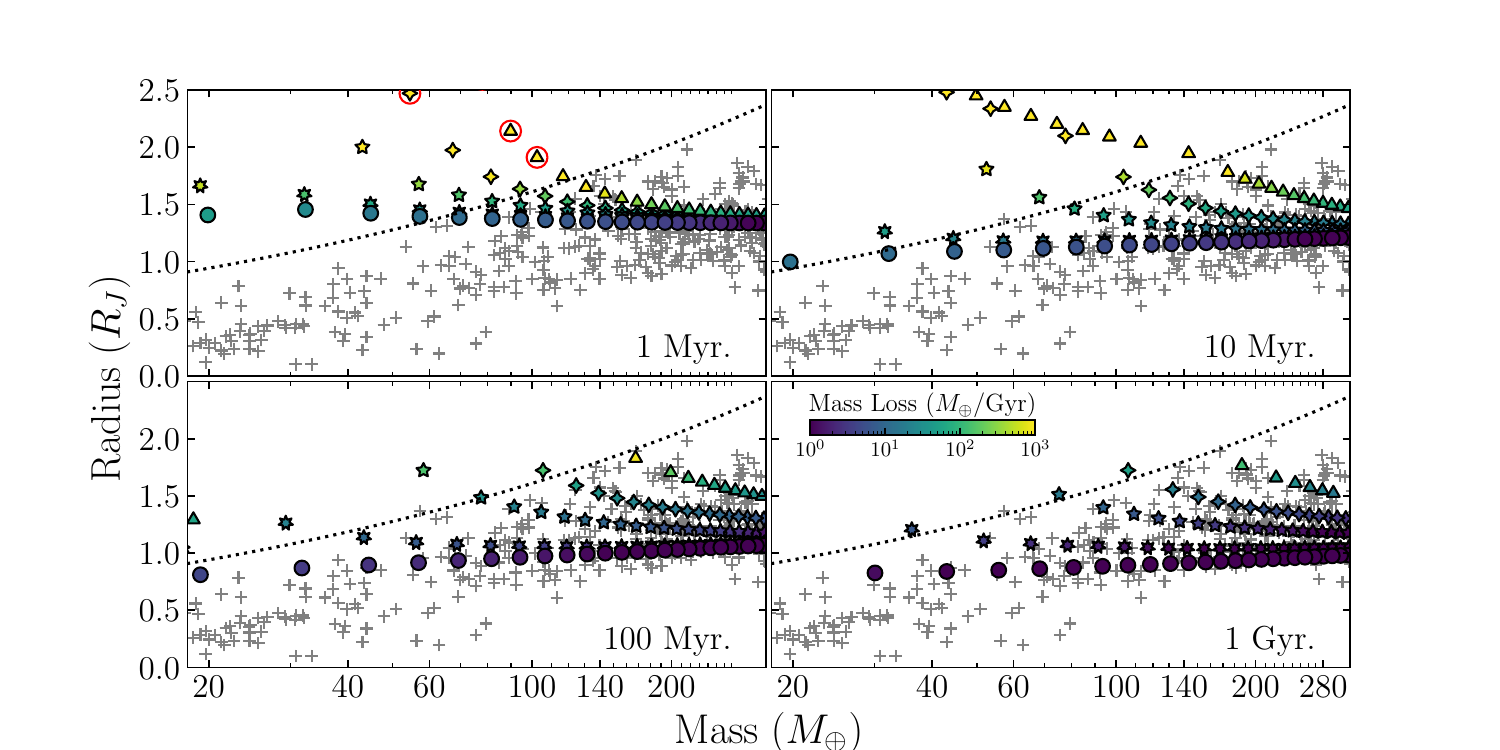}
    \caption{A grid of planets over a range of masses and semi-major axes undergoing mass loss and thermal evolution, color-coded with respect to their rate of mass loss, and different symbols corresponding to semimajor axes (triangles to circles representing 0.03, 0.04, 0.05, 0.07, and 0.1 AU). Planets circled in red have overflowed their Roche lobes.  Observed exoplanets are plotted beneath in grey, with the constant density $0.1 \gcc$ boundary shown as a dotted line.  Runaway mass loss rapidly strips low-density planets down to their cores so that within $\sim$100 Myr, the region above the density boundary is largely clear of planets.  Some additional planets near the boundary continue to lose mass, and a few undergo runaway mass loss later on (for example, the 0.04 AU planet above the density boundary at 1 Gyr will be lost within the next $\sim$ 250 Myr).  An animated version of this figure is available in the ancillary files which shows mass vs. radius and colored by the mass loss rate from 1 Myr to 3 Gyr over 14 seconds instead of time slices.}
    \label{fig:massLossAge}
\end{figure*}

\section{Results}\label{sec:results}
Fig.~\ref{fig:exampleTracks} illustrates our radius and mass evolution tracks.  We show three representative cases for a Saturn-mass planet.  At Saturn's orbit (9.58 AU), the planet experiences no mass loss and cools normally.  At 0.07 AU, the planet is a fairly typical hot Saturn, which cools to an inflated equilibrium where the luminosity equals the anomalous heating \citep[see e.g.][]{Thorngren2021}.  The planet experiences modest mass loss over the course of its lifetime.

Placing a Saturn at 0.05 AU has a very different outcome.  Although the planet does not initially overflow its Roche lobe nor is the initial mass-loss rate particularly high, the slow erosion of mass accelerates into the catastrophic removal of the envelope at $\sim$400 Myrs. This run-away mass loss is effected by an adiabatic expansion of the planet as it enlarges while losing mass at constant specific entropy.  Critically, the rate of mass loss scales as the radius cubed (recall Eq. \ref{eq:massLoss}), enhanced further by the tidal factor $K_t$.  Furthermore, at very low densities the mass-radius relationship becomes particularly steep (see Fig, \ref{fig:overview}).  The result is a positive feedback loop that reaches a head at around 100 Myr, when the planet grows in size rapidly, overflows its Roche lobe, rapidly losing mass and expanding the semi-major axis.  Over the course of ~30 Myr., the planet goes from 58 to 17.4 $M_\oplus$ and moves out to 0.95 AU.  What's left of the envelope continues to be stripped away by XUV-driven mass-loss, leaving less than 1 $M_\oplus$ of H/He on top of the 10.67 $M_\oplus$ core by 1 Gyr.

There is a subtle feature of this process which \citet{Valsecchi2014} observed and which we reproduce.  After Roche Lobe overflow starts, the orbit quickly expands just far enough to shut it off again (sacrificing some mass in the process).  However, unlike in \cite{Valsecchi2014}, XUV-mass loss is still taking place, and again expands the radius into the overflow regime.  Thus the planet is continuously very close to filling its Roche lobe and periodically exceeding it.  We do not explore this process in great detail here, as for our purposes it is enough to know that the process is very fast compared to thermal evolution, and that the growth in the semimajor axes is directly proportional to the mass lost.

Our result of runaway mass loss is reminiscent of what was reported by \citet{Kurokawa2014} which has since then been criticized for their use of fixed $\eta$ and overestimating the mass loss rate by fixing the pressure at the bottom of the photoevaporative flow at 1 nanobar \citep{OwenLai2018, Ionov2018}. The latter authors did not find any evidence of runaway mass loss. Our study differs from \citet{Kurokawa2014} in that we rigorously account for the variable $\eta$, but it was mainly our treatment of the thermal evolution that allowed us to recover catastrophic erosion.  Compared with e.g. \cite{OwenLai2018}, we include the effect of anomalous heating throughout the planet's thermal evolution instead of at the very end, which keeps our planets puffy. We have verified that when this anomalous heating is set to zero, the mass loss never reaches a runaway state.

A related issue we encounter is that during runaway, the gravitational potential at the planet's atmosphere drops below what is modeled by \citet{Caldiroli2022}, necessitating extrapolation.  The actual efficiency $\eta$ at low masses may be somewhat smaller than what we have adopted here \citep[see, e.g.][]{Ionov2018,Shematovich2014}. In the next section we will see that fortunately, our results are robust against even very small values of $\eta$.

\subsection{Carving out the Edges} \label{sec:population}
We are particularly interested in determining the overall effect of mass loss, especially the runaway variety, on the hot Saturn population, irrespective of particular initial distribution of host Saturns arising  from planet formation, which is outside the scope of this paper.  Instead, we seek to answer the more narrow question: if a hot Saturn of some mass is placed at a particular semimajor axis, can it survive?

We have run our mass loss model on a grid of masses and semimajor axes.  The masses were 15 values log-uniformly spaced from $20 M_\oplus$ to $1.5 M_J$ and the semimajor axes were 0.03, 0.04, 0.05, 0.07, and 0.1 AU.  The results are shown for several slices in time on Fig. \ref{fig:massLossAge} -- an animation is also available.  We assume a solar-mass star with a median amount of XUV irradiation (i.e., median spin) for its age.  Planets on short orbits experience more anomalous heating and consequently maintain a larger radius than those at longer periods after they cool.  Short-period, low-density planets are particularly susceptible to runaway mass loss and we observe a clear boundary at $\sim0.1\gcc$ within $\sim$100 Myr which persists to $\sim$1 Gyr, matching the observed density boundary of hot Saturns in radius-mass space.

\begin{figure*}
    \centering
    \includegraphics[width=\textwidth]{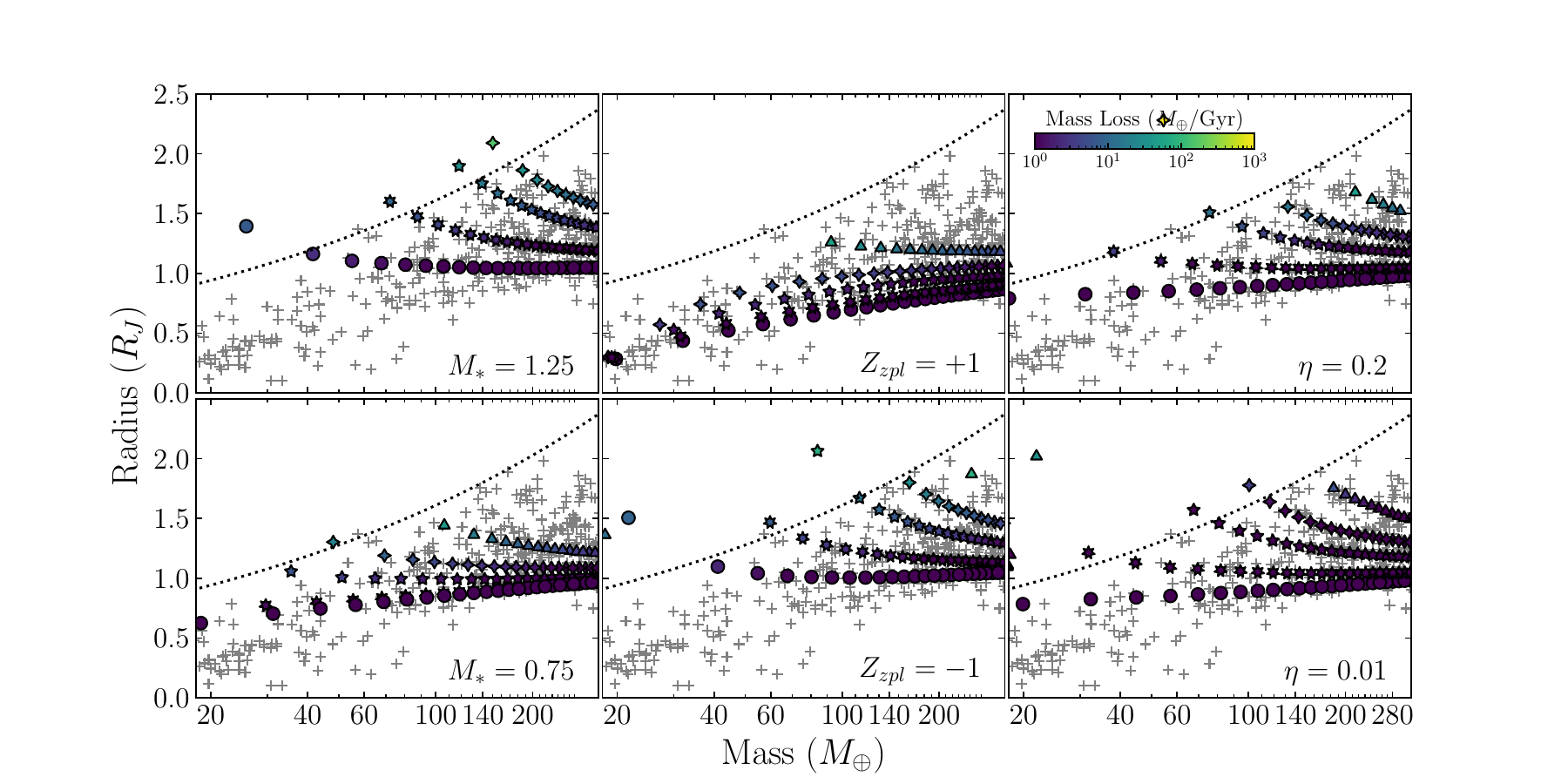}
    \caption{The model planets at 1 Gyr, varying the stellar mass (left), bulk metallicity (center), and XUV mass loss efficiency (right), color-coded with respect to the rate of mass loss. Symbols and the dotted line follow the convention of Figure \ref{fig:massLossAge}. For the metallicity $Z_{\rm pl}$, we take the $\pm 1\sigma$ value away from the mean of the mass-metallicity distribution reported by \citet{Thorngren2018}.  This range is broad; for example, a Saturn-mass planet with $-1\sigma$ metallicity has $15M_\oplus$ of metal, whereas at $+1\sigma$ it has $50M_\oplus$.  With all else equal, planets appear larger around more massive (and brighter) parent stars or for those with low bulk metallicity. These planets are more susceptible to mass loss and so the closest-in planets (triangles) are lost by 1 Gyr. At larger orbital distances, these puffier planets still survive owing to lower XUV flux and fill up the high mass ($\gtrsim$140$M_\oplus$) and large radii ($\gtrsim$1.5$R_J$). The right column shows that our results are only modestly sensitive to mass-loss efficiency with the region to the left of the density boundary being largely vacated between low and high $\eta$.}
    \label{fig:parameterStudy}
\end{figure*}

We have also considered the effect of stellar mass, the planet metallicity, and the effective XUV flux on the mass loss rates (Fig.~\ref{fig:parameterStudy}).  More massive stars have higher bolometric and XUV luminosities so planets around them are more inflated for a given mass and given orbital distance and therefore more susceptible to mass loss. The net result is that the closest-in planets (0.03 AU, triangle points) around higher mass stars lose their envelopes and lose their hot Saturn status while those at longer orbital distances survive against mass loss but stay puffy. Similarly, planets of lower bulk metallicity also appear puffy (due to their lower densities) and so the closest-in planets are lost to mass loss (note the lack of triangle points in the bottom center panel). Those at wider separations are stable against mass loss but will be larger than planets of higher bulk metallicity.

In Figure \ref{fig:massLossAge}, we see excess of planets of mass $\gtrsim$140$M_\oplus$ with radii above 1.5 $R_J$ in the observations compared to our fiducial model. Our parameter study (Figure \ref{fig:parameterStudy}) suggests that such planets are likely around more massive stars and/or they have lower bulk metallicity. Indeed, planets in our sample above 1.5 $R_J$ all orbit around stars more massive than the Sun, with an average mass of 1.41 $M_\odot$ (vs. $\sim1.1  M_\odot$ for $1R_J < R < 1.5 R_J$), corroborating our results.

The right panels of Fig.~\ref{fig:parameterStudy} illustrate the effect of varying $\eta$. As expected, closest-in planets are lost for higher fixed mass loss efficiency (note the lack of triangle points in the top right panel). Even at a low $\eta$=0.01, the low-density desert is largely cleared out by 1 Gyr, demonstrating that $\sim0.1\gcc$ is the boundary for a runaway mass loss (i.e., when the planet enters the runaway phase, they will be stripped of their sub-Saturn status regardless of the details of how they arrived to such a boundary).

\begin{figure*}
    \centering
    \includegraphics[width=\textwidth]{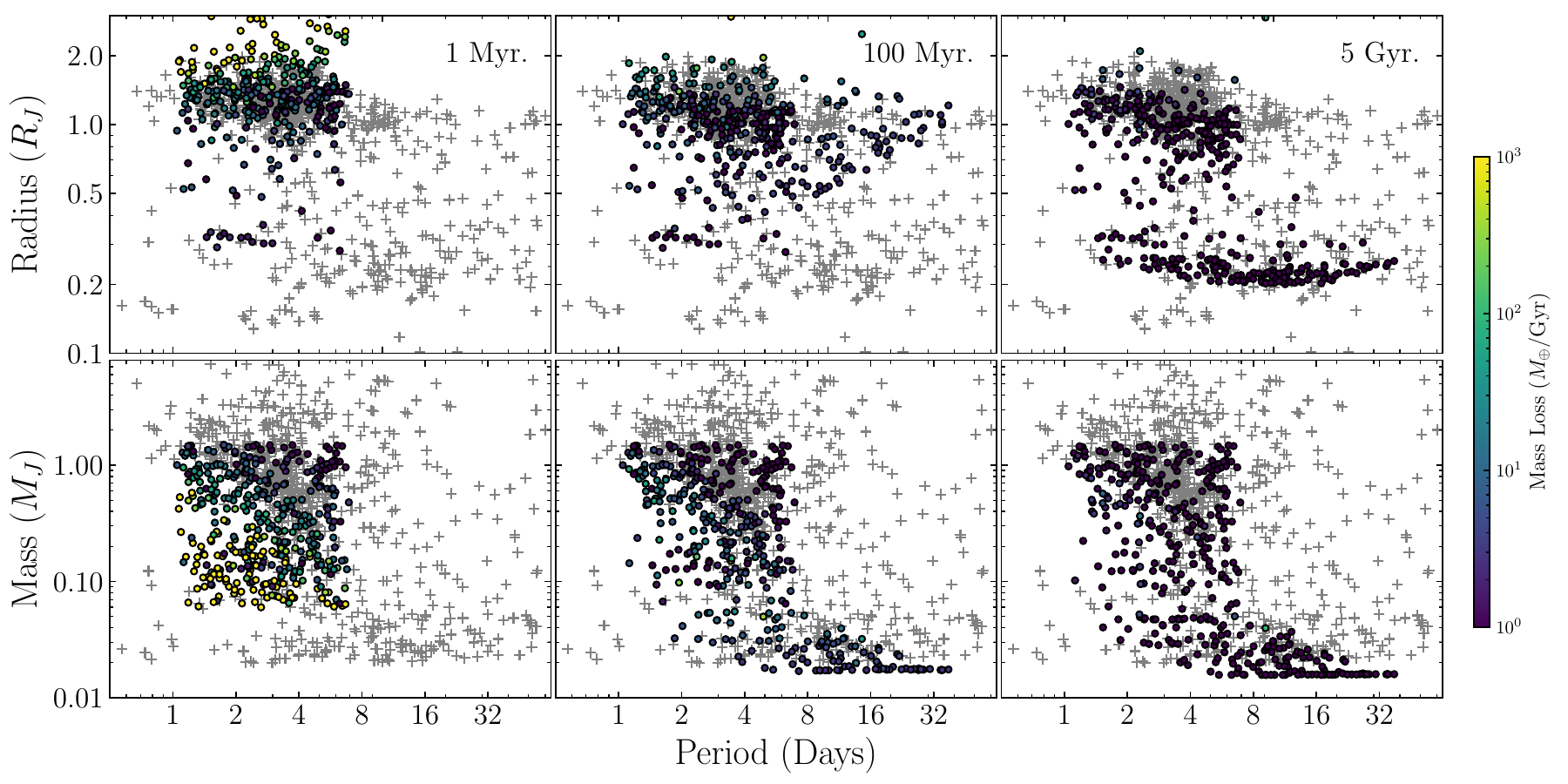}
    \caption{Model planets with random initial properties, plotted as radius against period (top) and mass against period (bottom) at 1 Myr, 100 Myr, and 5 Gyr.  The color indicates the rate of mass loss, and the observed population is plotted in grey.  In radius-period space, runaway mass loss carves out the upper edge of the radius distribution, removing the planets $\gtrsim$2$R_J$ from 1 Myr transforming them to $\lesssim$0.3$R_J$ by 5 Gyr. In mass-period space, runaway mass loss removes the lower left half of the triangle (the yellow points) at 1 Myr and vacate that region by 5 Gyr, transforming them into $\lesssim$0.04$M_J$ planets.  An animated version of this figure is available in the ancillary files which shows radius vs period and mass vs period and colored by the mass loss rate from 1 Myr to 5 Gyr over 14 seconds instead of time slices.} \label{fig:randomTracks}
\end{figure*}

We now explore how the runaway mass loss manifests in the mass-period and radius-period spaces.  First, we draw 400 sets of mass, semimajor axis, stellar mass, and metallicity, distributed log-uniformly from 20 to $1.5 M_J$ in mass, uniformly from 0.8 to 1.2 $M_\odot$ in stellar mass, log-uniformly from 0.03 AU to 0.1 AU in semimajor-axis, and the metallicity according to the mass-metallicity relation \citep{Thorngren2021}. These planets are evolved out to 5 Gyr and the results are shown in Fig.~\ref{fig:randomTracks}.  We observe that most of the low-mass, short-period planets have large radii and are quickly reduced to stripped cores.  All model planets near or above $2R_J$ are destroyed by runaway inflation within 100 Myr, shaping the upper edge of the radius distribution. Over 5 Gyr, the sub-Jovian desert in mass-period space is largely carved out by mass loss although it is not fully cleared, mostly due to high-bulk metallicity planets which have small radii, suggesting the formation of hot Saturns disfavors those with high densities.  In some cases, tidal inspiral may lead to runaway mass-loss, though careful calculation would be needed to examine this possibility as the XUV luminosity will have dropped significantly on inspiral timescales ($\sim$Gyr).

Another interesting feature of our evolution models is that the planet population appears to pass through a few phases in time.  In the first few tens of Myr, many planets are actively losing significant amounts of mass while undergoing orbital expansion; they have not reached their final mass or size yet.  At a few hundred Myr, the majority of planets have lost most of the mass that they are going to lose during their lifetime, but their envelopes are still too hot and remain puffy, giving rise to a large number of puffy planets ($R > 0.8 R_J$) even at relatively long periods and low mass (see Fig. \ref{fig:randomTracks}, middle column).  By around 500 Myr, most planets complete the runaway mass loss, though some continue to runaway late.  The population of early-runaway planets has cooled to below 0.5 $R_J$ and will continue to cool towards the evolved radius of around 0.2-0.3 $R_J$ over the subsequent Gyrs.  The existence of young puffy planets ($\sim$100 Myr, $\sim$0.5$R_J$, $\lesssim$0.03$M_J$) beyond $\sim$4 days is a unique prediction of our model that could be tested by e.g., TESS and follow-up mass measurements \citep[such as e.g.][]{Newton2019,Zhou2021}.

From our simple initial random distribution of planets, we find that the ones that survive against mass loss to fill in the sub-Jovian desert in mass-period and radius-period spaces are those that have high bulk metallicity (and so higher density). Our result agrees with the discovery of dense Neptune to Saturn size planets found within the desert such as LTT 9799 b \citep{Jenkins2020} and TOI-849 b \citep{Armstrong2020}, although these two planets could also be stripped massive cores of even larger giants.

\section{Discussion and Conclusion}\label{sec:disc}
Our primary conclusion is that hot Saturns can undergo runaway mass loss caused by positive feedback between the adiabatic expansion of the radius and the mass loss (Fig.~\ref{fig:exampleTracks}).  In many cases, this leads to Roche lobe overflow, though this is not necessary to remove the planets' envelope.  Over a range of stellar masses, planet bulk metallicity, and the XUV mass loss efficiency, we find that this process can robustly sculpt the low density boundary at $\sim0.1\gcc$ (Figs.~\ref{fig:massLossAge} and \ref{fig:parameterStudy}).

In radius-period space, the radii for our initial distribution already show a relative paucity of planets at $\sim$0.3--1.0$R_J$ inside $\sim$3 days, similar to what is observed.  This initial radius-period paucity arises from a sudden rise in radius over a small dynamic range in mass at high incident flux (see the model mass-radius relation curves in the bottom panel of Figure \ref{fig:overview}). Subsequent XUV-driven mass-loss quickly removes planets from the sub-Jovian desert in mass-period space, though somewhat too many planets appear to remain in the desert compared to the observations (Fig.~\ref{fig:randomTracks}).  To completely clear out the desert, we may need to more carefully account for Roche lobe overflow and tidal processes \citep[e.g.][]{Valsecchi2015} in our models, but perhaps more important is the relatively simple choice of input distribution, especially on planet mass and period.  For example, \citet{Hallatt2022} were able to reproduce the desert (in terms of the decline in sub-Saturn occurrence rate at short periods) through XUV-driven mass-loss and a careful consideration of the initial core mass function and gas accretion theory. It also remains possible that some of the short period giants arrived there via high-eccentricity migration which contribute to shaping the upper edge of the desert in the mass-period space \citep{Matsakos2016,OwenLai2018}.

Planets that lose their envelopes collect around $\sim$0.2 to $\sim$0.3 $R_J$ based on their initial core mass, which we imposed in Section \ref{sec:methods}.  This assumption does not change the triggering of runaway mass-loss, but it does impact the final state.  This suggests that if we are able to identify probable stripped cores, they would be a unique window into the deep interiors of giant planets.  A few such candidates have already been proposed \citep{Petigura2017,Jenkins2020,Armstrong2020}, and a targeted study would likely identify more.

Our results differ somewhat from \citet{OwenLai2018} and \cite{Ionov2018}, who argue that photoevaporative mass loss is unable to explain the upper edge of the sub-Jovian desert (i.e., hot Saturns are resilient against mass loss).  First, we argue that the effect of mass loss is more clearly seen in radius-mass space rather than mass-period or radius-period. Second, and more importantly, we include the anomalous heating of hot giants found in \citet{Thorngren2018} throughout the planets' evolution, resulting in larger planets more vulnerable to mass loss than \citet{OwenLai2018}, who evolve their planets without heating and then inflate them after mass-loss has completed.  We believe our approach is more consistent with the observations, as in \cite{Thorngren2021}, it was found that young planets are as large or larger than their older equivalents once the present-day incident flux is controlled for.

Recently, \cite{Vissapragada2022} examined a set of seven transiting planets on the (mostly upper) edge of the desert in mass-period space.  They found that none of the planets were undergoing massive mass loss despite sizable XUV fluxes.  We do not find their result to be in tension with ours as all the planets they investigated have densities larger than the 0.1 ${\rm g\,cm^{-3}}$ boundary, ranging from 0.172 to 0.717 ${\rm g\,cm^{-3}}$.  The lowest density planet, WASP-177 b is still interesting as a planet near the density edge, but they find that its mass loss rate is quite low, which could be due to its old age of $\sim10$ Gyr \citep{Turner2019}, as XUV luminosity declines as the star ages.

Future work on this area could investigate the potential mass-loss histories of giant planets near the low-density boundary.  One good candidate for this is WASP-17 b \citep{Anderson2010}, the largest known planet with a well-determined mass and radius, and one which is likely to be losing mass at a prodigious rate.  Qatar-10 b \citep{Alsubai2019} and WASP-76 b \citep{West2016} are also good cases for substantial ongoing mass loss.  Also of interest are the smaller and lower-mass planets which are nevertheless near the 0.1 ${\rm g\,cm^{-3}}$ boundary, including WASP-20 A b \citep{Anderson2015} and WASP-153 b \citep{Demangeon2018}. While these planets are not experiencing runaway mass-loss at the present day, they likely are experiencing some loss, and they may be nearing runaway.  Measuring both the mass loss rates and XUV fluxes of their parent stars could help to pinpoint some of the fine details of this process.

\acknowledgments{We thank Andrea Caldiroli, Jonathan Fortney, Ruth Murray-Clay, and Shreyas Vissapragada for helpful discussions, and the anonymous reviewer for their thoughtful suggestions and comments. D.P.T thanks the Trottier Institute for Research on Exoplanets (iREx) for support via the Trottier Fellowship and Johns Hopkins University for support via the Davis Fellowship. E.J.L. gratefully acknowledges support from NSERC, FRQNT, the Trottier Space Institute, and the William Dawson Scholarship from McGill.  E.D.L. would like to acknowledge support from the GSFC Sellers Exoplanet Environments Collaboration (SEEC), which is funded in part by the NASA Planetary Science Division’s Internal Scientist Funding Model.}

\bibliography{bibliography}
\end{document}